\documentclass[aps,twocolumn,prl]{revtex4-1}

\usepackage{graphicx}% Include figure files
\usepackage{dcolumn}% Align table columns on decimal point
\usepackage{bm}% bold math
\usepackage{xcolor}
\usepackage{amssymb}   % for math
\usepackage{amsmath}
\usepackage{ulem}

\begin{document}
	
\title{Majorana quantization and  half-integer thermal quantum Hall effect \\in a Kitaev spin liquid}
	
%\title{Majorana quantization of thermal Hall conductance in a Kitaev quantum magnet }

%\title{ Discovery of half-integer quantum thermal Hall effect in a Kitaev quantum spin liquid}

\author{Y.\,Kasahara$^1$}
\author{T.\,Ohnishi$^1$}
\author{N.\,Kurita$^2$}
\author{H.\,Tanaka$^2$}
\author{J.\,Nasu$^2$}
\author{Y.\,Motome$^3$}
\author{T.\,Shibauchi$^4$}
\author{Y.\,Matsuda$^1$}

\affiliation{$^1$Department of Physics, Kyoto University, Kyoto 606-8502, Japan}
\affiliation{$^2$Department of Physics, Tokyo Institute of Technology, Meguro, Tokyo 152-8551, Japan}
\affiliation{$^3$Department of Applied Physics, University of Tokyo, Bunkyo, Tokyo 113-8656, Japan}
\affiliation{$^4$Department of Advanced Materials Science, University of Tokyo, Chiba 277-8561, Japan}

\maketitle

{\bf 
The quantum Hall effect (QHE) in two-dimensional (2D) electron gases, which is one of the most striking phenomena in condensed matter physics,  involves the topologically protected dissipationless charge current flow along the edges of the sample. Integer or fractional electrical conductance are measured  in units of $e^2/2\pi \hbar$, which is associated with  edge currents of electrons or quasiparticles with fractional charges, respectively.   Here we discover a novel type of quantization of the Hall effect  in an insulating 2D quantum magnet \cite{Kitaev06}.   In $\alpha$-RuCl$_3$ with dominant Kitaev interaction on 2D honeycomb lattice \cite{Jackeli09,Trebst17,Kim15,Banerjee16,Sandilands15,Nasu16}, the application of a parallel magnetic field destroys  the long-range magnetic order, leading to a field-induced quantum spin liquid (QSL) ground state with massive entanglement of local spins \cite{Yadav16,Baek,Wolter,Leahy16,Hentrich}.  In the low-temperature regime of the QSL state,  we report that the 2D thermal Hall conductance $\kappa_{xy}^{\rm 2D}$ reaches a quantum plateau as a function of  applied magnetic field.  That is,  $\kappa_{xy}^{\rm 2D}/T$ attains a quantization value of $(\pi/12)(k_B^2/\hbar)$, which is exactly half of $\kappa_{xy}^{\rm 2D}/T$ in the integer QHE.  This half-integer thermal Hall conductance observed in a bulk material  is a direct signature of topologically protected chiral edge currents of charge neutral Majorana fermions, particles that are their own antiparticles, which possess half degrees of freedom of conventional fermions  \cite{Read00,Sumiyoshi13,Nomura12,Nasu17}.  These signatures  demonstrate  the fractionalization of spins into itinerant Majorana fermions and $Z_2$ fluxes predicted in a Kitaev QSL \cite{Kitaev06,Trebst17}.   Above a critical magnetic field, the quantization disappears and  $\kappa_{xy}^{\rm 2D}/T$ goes to zero rapidly, indicating a topological quantum phase transition between the states with and without chiral Majorana edge modes.  Emergent Majorana fermions in a quantum magnet  are expected to have a major impact on strongly correlated topological quantum matter.
}

The topological states of matter are described in terms of topological invariant quantities whose values are quantized.  The most popular quantity to prove these states is the electrical Hall conductivity.  In the quantum Hall state,  the Hall conductance $\sigma_{xy}^{\rm 2D}$ is quantized in units of $e^2/2\pi\hbar$ as $\sigma_{xy}^{\rm 2D}=q(e^2/2\pi\hbar)$, where $q$ is integer in integer QHE and is fraction which, with very few exceptions, has an odd denominator  in fractional QHE.  These quantizations attest to topologically ordered states.  Another topological invariant in the topological phase is the 2D thermal Hall conductance; thermal Hall conductivity per 2D sheet $\kappa_{xy}^{\rm 2D}$ is quantized in units of $(\pi/6)(k_B^2/\hbar)T$ as
\begin{equation}
\kappa_{xy}^{\rm 2D}/T=q(\pi/6)(k_B^2/\hbar).
\end{equation}
 Although thermal Hall conductivity is much harder to measure than electrical Hall conductivity, it has a clear advantage in revealing the topological phases possessing charge neutral excitations that cannot be detected by the electrical Hall conductivity.   In particular, $q=1/2$ state with positive thermal Hall sign is a decisive manifestation of the charge neutral edge currents of Majorana particles (Figs.\,1a and 1b),  distinguishing unambiguously between the different candidate topological orders.  We note that a Majorana quantized phase characterized by $q=1/2$ has been predicted in chiral topological superconductors \cite{Read00,Sumiyoshi13,Nomura12}. However, as the topological superconductivity in bulk materials have not been fully established, previous experiments searching for Majorana fermions have focused on the proximity effect between conventional superconductors and nanowires or topological materials \cite{Mourik,Nadj,Das12,He17}.  Here we present a fundamentally different approach to this issue and perform direct measurements of the thermal Hall conductance in a bulk insulating magnet.

 \begin{figure}[t]
	\begin{center}
		%\vspace*{-20mm}
		%\hspace*{5mm}
		\includegraphics[width=1.0\linewidth]{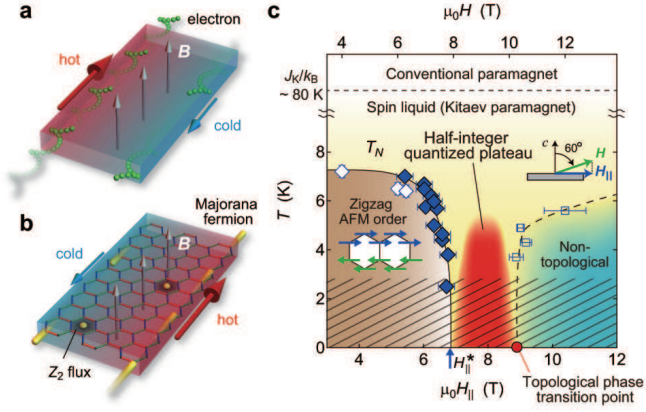}
		%\vspace{10mm}
		\caption{
			{\bf Chiral Majorana edge currents and temperature-magnetic field phase diagram of $\alpha$-RuCl$_3$.} {\bf a,b,} Schematic illustrations of heat conductions in the integer quantum Hall state of 2D electron gas ({\bf a}) and Kitaev QSL state ({\bf b}) in magnetic field applied perpendicular to the planes (gray arrows).  In the red (blue) regime, the temperature is higher (lower). The red and blue arrows represent thermal flow. In the quantum Hall state, the skipping orbits of electrons (green spheres) at the edge, which form 1D edge channels, conduct heat and $\kappa_{xy}$ is negative in sign.    In the Kitaev QSL state,  spins are fractionalized into Majorana fermions (yellow spheres) and $Z_2$ fluxes (black hexagons).  The heat is carried by chiral  edge currents of charge neutral Majorana fermions  and $\kappa_{xy}$ is positive in sign.	{\bf c,} Phase diagram of $\alpha$-RuCl$_3$ in tilted field of $\theta=60^\circ$. Open and closed diamonds represent the onset temperature of AFM order with zigzag type $T_N$ determined by $T$- and $H$-dependences of $\kappa_{xx}$, respectively. Below $T\sim J_K/k_B$, the spin liquid (Kitaev paramagnetic) state appears. At $\mu_0H_\parallel^\ast\sim7$\,T,   $T_N$ vanishes (blue arrow).   A half-integer quantized plateau of 2D thermal Hall conductance is observed in the red regime.  
Open blue squares represent the fields at which the thermal Hall response disappears.  Red circle indicates a topological phase transition point that separates  the non-trivial QSL state with  topologically protected chiral Majorana edge currents and a  trivial state, such as non-topological spin liquid or forced ferromagnetic state.}
	\end{center}
	%	\vspace{-5mm}
\end{figure}

The systems composed of interacting 1/2 spins on a honeycomb lattice with bond-directional exchange interactions $J_K$ are of vital interest, as they host QSL ground states where topological excitations emerge \cite{Kitaev06}.   This Kitaev QSL exhibits two types of fractionalized quasiparticle excitations, i.e. itinerant (mobile) Majorana fermions and   $Z_2$ fluxes with a gap.   The Majorana fermion  has a massless (gapless) Dirac-type dispersion in zero field.  In magnetic fields, a novel Majorana fermion system, which is characterized by the bulk gap and  gapless edge modes,  is realized \cite{Kitaev06,Trebst17}, and the $Z_2$ flux obeys anyonic statistics. 
  
  Recently, a strongly spin-orbit coupled  Mott insulator $\alpha$-RuCl$_3$  has emerged as a prime candidate for hosting an approximate Kitaev QSL.   In this compound, local  $j_\mathrm{eff}=1/2$ pseudospins  are aligned in 2D honeycomb layer and the Kitaev interaction $J_K/k_B \sim 80$\,K is predominant \cite{Banerjee16,Sandilands15,Nasu16}.     The system is in a spin-liquid (Kitaev paramagnetic) state below $\sim J_K/k_B$, and shows antiferromagnetic (AFM) order with zigzag spin structure (Fig.\,1c) at $T_N \approx7$\,K  \cite{Johnson15} due to non-Kitaev interactions, such as Heisenberg exchange and off-diagonal interactions.   Although the thermal Hall conductance has been measured in $\alpha$-RuCl$_3$, the quantization is not observed because the low temperature properties of the liquid state is masked by the AFM order \cite{Kasahara17}.

The  response of $\alpha$-RuCl$_3$ to magnetic fields is highly anisotropic \cite{Yadav16,Leahy16,Hentrich,Majumder,Chaloupka16}.  It has been reported that while $T_N$ is little influenced by external magnetic field perpendicular to the 2D plane,  $T_N$ is dramatically suppressed by the parallel field.    This highly anisotropic response is  confirmed by the measurements of longitudinal  thermal conductivity $\kappa_{xx}$ in  magnetic field {\boldmath $H$}  applied along various directions in the $ac$ plane as shown in the inset of Fig.\,2a,  where $H_{\parallel}=H\sin \theta$ and $H_{\perp}=H\cos\theta$ are the field component parallel and perpendicular to the $a$ axis, respectively, and $\theta$ is the angle between {\boldmath $H$} and the $c$ axis.    In zero field,  $\kappa_{xx}$ exhibits a distinct kink at $T_N$, as shown in Fig.\,2a.    While this kink  is observed in perpendicular field ($\theta=0^\circ$) of 12\,T at the same temperature,  no kink anomaly is observed in parallel field ($\theta=90^{\circ}$) of 7\,T \cite{Leahy16,Hentrich}.    In Fig.\,2a, we also plot $\kappa_{xx}$ in applied magnetic field of 8\,T  tilted away from the $c$ axis ($\theta=60^{\circ}$, $H_\parallel\sim7$\,T).      Similar to the case of parallel field, no kink anomaly is observed.   Figure\,1b displays the phase diagram in tilted field of $\theta=60^{\circ}$, where $T_N$ is plotted as a function of  $H_{\parallel}$.   We determined $T_N$ by the kink of $T$-dependence of $\kappa_{xx}$ and by the minimum in the $H$-dependence of $\kappa_{xx}$ (see Fig. 2b and Extended Data Figs.\,1 and 2).   The inset of Fig.\,2b shows $T_N$ plotted  as a function of $H_{\parallel}$ for $\theta=45^{\circ}, 60^{\circ}$ and $90^{\circ}$.   While $T_N$ for $\theta=60^{\circ}$  well coincides  with that for 90$^{\circ}$ and vanishes at the same critical field of $H_{\parallel}^*\approx 7$\,T,  $T_N$ for $45^{\circ}$ vanishes at around $H_{\parallel}\approx 6$\,T.   Although $T_N$ is not perfectly scaled by $H_{\parallel}$, these results demonstrate the quasi-2D nature of the magnetic properties.

 \begin{figure}[t]
	\begin{center}
		%\vspace*{-20mm}
		%\hspace*{5mm}
		\includegraphics[width=1.0\linewidth]{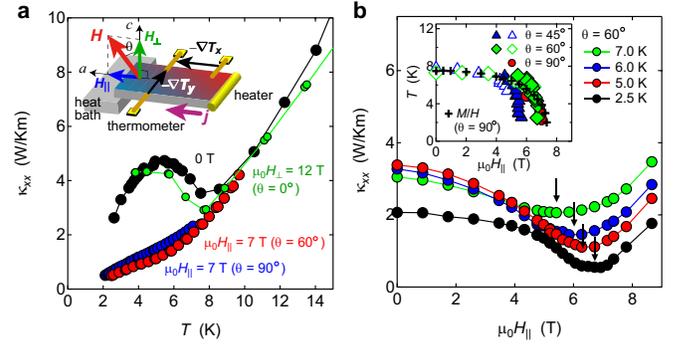}
		%\vspace{10mm}
		\caption{
			{\bf Longitudinal thermal conductivity in $\alpha$-RuCl$_3$.} {\bf a,} Temperature dependence of $\kappa_{xx}$ in magnetic field \bm{$H$} applied along various directions in the $ac$ plane. Inset illustrates a schematic of the measurement setup for $\kappa_{xx}$ and $\kappa_{xy}$. {\bf b,}  $\kappa_{xx}$ at  $\theta$=60$^{\circ}$ plotted as a function of parallel field component $H_{\parallel}$.  Inset shows $T_N$ vs. $H_{\parallel}$ at different field directions.  $T_N$ is determined by the $T$-dependence of $\kappa_{xx}$ shown in Fig.\,2a (open symbols) and by the minimum in the $H$-dependence of $\kappa_{xx}$ (filled symbols) shown by arrows in the main panel.  The crosses are $T_N$ for $\theta=90^\circ$ determined from magnetic susceptibility measurements \cite{Banerjee17}. }
	\end{center}
	%	\vspace{-5mm}
\end{figure}

Above $H_{\parallel}^\ast$ where the AFM order melts,   the presence of a peculiar spin liquid state has been suggested by the nuclear magnetic resonance (NMR) and neutron scattering measurements.  The former  reports  the presence of  spin gap \cite{Jansa} and the latter  reveals unusual  continuous spin excitations \cite{Banerjee17}.  These magnetic properties are consistent with those expected in a Kitaev-type spin liquid state.    %The most fascinating question is whether this field-induced state is in a Kitaev QSL state characterized by the chiral Majorana edge currents. 
  
  To study the thermal Hall effect in the spin liquid state above $H_{\parallel}^\ast$,  $\kappa_{xy}$ is measured by sweeping field in tilted directions and  obtained by anti-symmetrizing thermal response with respect to the field direction.  In this configuration,  Hall response is determined by $H_{\perp}$.      Since the magnitude of $\kappa_{xy}$ is extremely small compared to  $\kappa_{xx}$ in $\alpha$-RuCl$_3$,  special care was taken to detect the intrinsic thermal Hall signal \cite{Watanabe16}.    The experimental error in determining $\kappa_{xy}$ caused by the uncertainty in measuring the distance between the Hall contacts and the thickness of the crystal is within $\pm 2$\%.     Figures\,3a, b, and c depict $\kappa_{xy}/T$ at $\theta=60^{\circ}$ plotted as a function of $H_{\perp}$ above $H_{\parallel}^\ast$ at low temperatures. Below 3.7\,K, the transverse thermal gradient is hard to detect within our resolution.

  In the AFM state, $\kappa_{xy}/T$ is extremely small (see Extended Data Fig.\,3).  Upon entering the field-induced spin liquid state,   $\kappa_{xy}/T$, which is positive in sign,  increases rapidly.   The most striking feature is that  $\kappa_{xy}/T$ attains a plateau in the field range of 4.5\,T$<\mu_0H_{\perp}<$4.8-5.0\,T, as displayed in Figs.\,3a, b and c.  The right axes represent $\kappa_{xy}^{\rm 2D}$ in units of quantum thermal Hall conductance $(\pi/6)(k_B^2/\hbar)T$, where $\kappa_{xy}^{\rm 2D}=\kappa_{xy}d$ with the layer distance $d=5.72$\,\AA\ \cite{Johnson15}.    Remarkably, the plateau is very close to the half of quantum thermal Hall conductance reported in the integer quantum Hall system \cite{MBanerjee17} within the error of 3\%, demonstrating the emergence of half-integer thermal Hall conductance plateau.   Above $\mu_0H_{\perp}\approx 5.0$\,T, $\kappa_{xy}^{\rm 2D}/T$ decreases rapidly and vanishes.   At 4.3 and 4.9\,K, slight increase of $\kappa_{xy}^{\rm 2D}/T$ is observed before the reduction, while it is absent at 3.7\,K. Although the plateau behaviour seems to be preserved at 5.6\,K, $\kappa_{xy}^{\rm 2D}/T$ slightly deviates from the quantized value.   At higher temperatures, the plateau behaviour disappears (Fig.\,3d).

 \begin{figure}[t]
	\begin{center}
		%\vspace*{-20mm}
		%\hspace*{5mm}
%		\includegraphics[width=1.0\linewidth]{Fig3.eps}
		\includegraphics[width=0.9\linewidth]{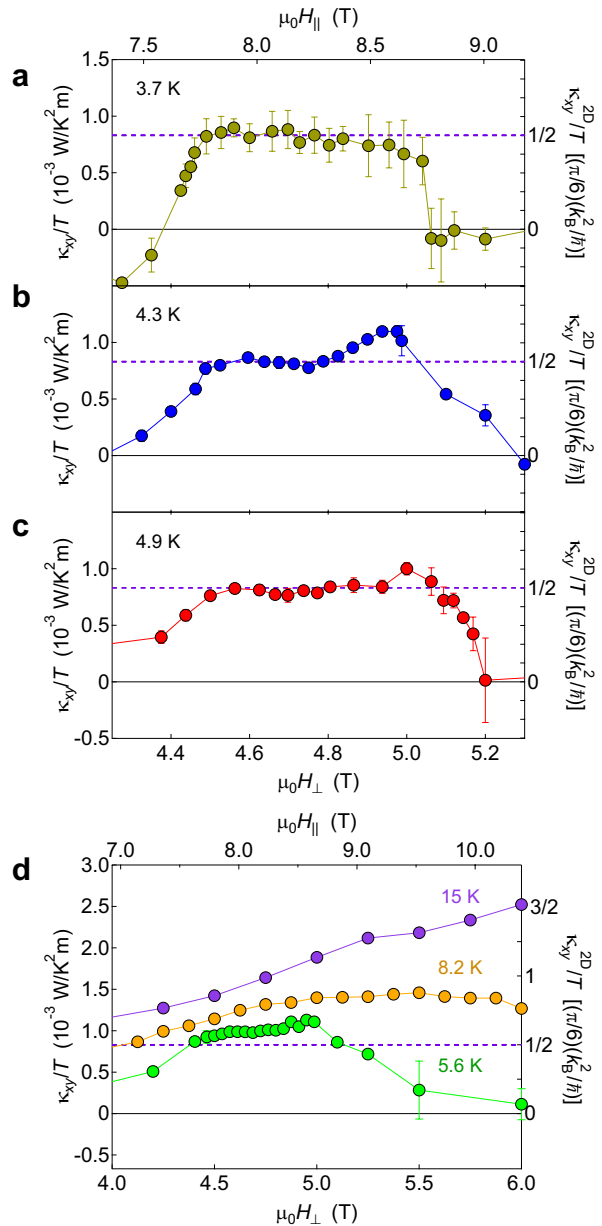}
%		\includegraphics[width=220mm, angle=90]{Fig3.eps}
		%\vspace{10mm}
		\caption{
			{\bf Half-integer thermal Hall conductance plateau.} {\bf a-d,} Thermal Hall conductivity $\kappa_{xy}/T$ in tilted field of $\theta=60^\circ$ plotted as a function of $H_\perp$. The top axes show the parallel field component $H_\parallel$. The right scales represent the 2D thermal Hall conductance $\kappa_{xy}^{\rm 2D}/T$ in units of $(\pi/6)(k_B^2/\hbar)$. Violet dashed lines represent the half-integer thermal Hall conductance, $\kappa_{xy}^{\rm 2D}/[T(\pi/6)(k_B^2/\hbar)]=1/2$.  } 
	\end{center}
	%	\vspace{-5mm}
\end{figure}

The temperature dependence of $\kappa_{xy}/T$ at the magnetic fields where the plateau is observed is shown in Fig.\,4.  The half-integer thermal Hall conductance is observable up to $\sim$5.5\,K, above which $\kappa_{xy}/T$ increases rapidly with $T$.   As shown in the inset of Fig.\,4, $\kappa_{xy}/T$ decreases after reaching  a maximum at around 15\,K and   nearly vanishes above $\sim$ 60\,K (see Extended Data Fig.\,4).  In usual Heisenberg systems, finite  thermal Hall effect can appear in spin-liquid states in the presence of  Dzyaloshinsky-Moriya interaction \cite{Han16}. However, such an interaction in $\alpha$-RuCl$_3$ is negligible as it is less than 5\% of $J_K$ \cite{Winter16}. Moreover, the phonon thermal Hall conductivity is three orders of magnitude smaller than the observed $\kappa_{xy}/T$ in the spin-liquid state and shows essentially different $T$-dependence \cite{Sugii17}.    As the vanishing temperature of $\kappa_{xy}/T$ is close to  Kitaev interaction, it is natural to consider that the finite thermal Hall signal reflects unusual quasiparticle excitations inherent to the spin liquid state governed by the Kitaev interaction.

 \begin{figure}[t]
	\begin{center}
		%\vspace*{-20mm}
		%\hspace*{5mm}
		\includegraphics[width=1.0\linewidth]{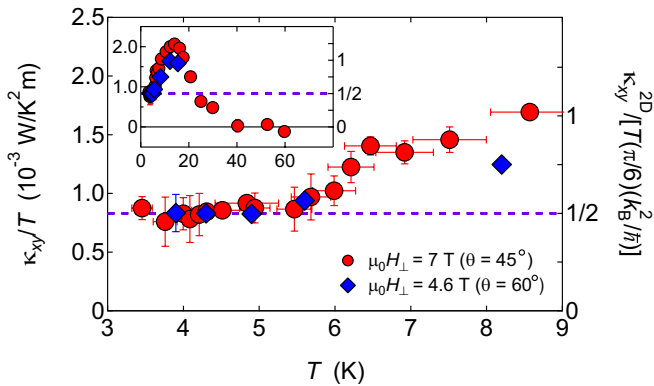}
		%\vspace{10mm}
		\caption{
			{\bf Temperature dependence of the thermal Hall conductance.} The main panel shows $\kappa_{xy}/T$ in tilted fields of $\theta=45^\circ$ and $60^\circ$ at  $\mu_0H_{\perp}=$7\,T and 4.6\,T, respectively, at which quantized thermal Hall conductance plateau is observed at low temperatures.  The right scale is the 2D thermal Hall conductance $\kappa_{xy}^{\rm 2D}/T$ in units of $(\pi/6)(k_B^2/\hbar)$. Violet dashed line represents the half-integer thermal Hall conductance, $\kappa_{xy}^{\rm 2D}/[T(\pi/6)(k_B^2/\hbar)]=1/2$.  Inset shows the same data in a wider temperature regime. }
	\end{center}
	%	\vspace{-5mm}
\end{figure}

In Eq.(1), the coefficient $q$ gives the chiral central charge of the gapless boundary  modes, which propagate along one direction.   Central charge represents  a number of freedom of 1D gapless modes; it is unity for conventional fermions, while it is 1/2 for Majorana fermions whose degrees of freedom is half of conventional fermions.  Integer quantum Hall system with the bulk Chern number $\nu$  has $\nu$  boundary modes with $q=\nu$, while  a Kitaev QSL with the Chern number $\nu$ has $\nu$ Majorana boundary modes with $q=\nu/2$. Thus the observed half-integer thermal Hall conductance provides direct evidence of the chiral Majorana edge currents.   We also note that the positive Hall sign is also consistent with that predicted in the Kitaev QSL \cite{Kitaev06}.  In pure Kitaev model,  the excitation energy of $Z_2$ flux is estimated to be  $\Delta_F/k_B\sim 0.06 J_K/k_B\sim 5.5$\,K \cite{Nasu16}.  Experimentally NMR reports the magnetic excitation gap of  $\sim$10\,K, which  is closely related to the  flux gap \cite{Hentrich,Jansa}.  The recent numerical results of the thermal Hall conductance for the 2D pure Kitaev model calculated by the quantum Monte Carlo method  show the quantization occurs slightly below $\Delta_F/k_B$ \cite{Nasu17}. Therefore,  the thermal Hall quantization which preserves up to $\sim 5$\,K is consistent with the  excitation gap.  

In the plateau regime of $\kappa_{xy}$,  no anomaly is observed in $\kappa_{xx}$.  This is likely because phonon contributions largely dominate over the fermionic excitations arising from spins in $\kappa_{xx}$ in the whole $T$-range \cite{Hirobe17,Yu17}.  Moreover, due to  the strong spin-phonon coupling in $\alpha$-RuCl$_3$ \cite{Leahy16},  the phonon conductivity is expected to show a complicated $H$-dependence.   The fact that $\kappa_{xy}$ vanishes at the highest fields as shown in Figs.\,3a-c  provides direct evidence that  the thermal Hall effect  is not  influenced by phonons, demonstrating that $\kappa_{xy}$ is a unique and  powerful probe in the search for Majorana quantization. 

We stress that the half-integer thermal Hall conductance in a bulk material is a direct consequence of the chiral Majorana edge current. Recent experiments based on the proximity effect between a quantum anomalous Hall insulator and a conventional superconductor have reported a signature of chiral Majorana edge modes \cite{He17}. However, this is based on the observation of half-integer quantization of the longitudinal electrical conductance via the scattering matrix effect between the edge states of the insulator and superconductor. Moreover, Majorana fermions in the Kitaev magnets and topological superconductors have  essentially different aspects. In the former, strong correlations give rise to the emergent Majorana fermions, while in the latter they do not play a role. In addition, Majorana fermions exist inside the bulk of a sample in the Kitaev QSL state, in sharp contrast to topological superconductors where they appear only at the edges. The distinct nature is presumably supported by the fact that the quantum plateau disappears below $\sim 400$\,mK in the topological superconductor device \cite{He17}, whereas it is preserved up to $\sim 5$\,K in $\alpha$-RuCl$_3$.

The numerical results of the thermal Hall effect for the 2D pure Kitaev model show that $\kappa_{xy}^{\rm 2D}/T$ does not exceed the half-quantized value at any temperature in the weak field regime.  The calculation shows that $\kappa_{xy}^{\rm 2D}/T$ is reduced from the quantized value as the temperature is raised \cite{Nasu17},  which is an opposite tendency to the experimental results shown in Fig.\,4 and its inset.  Moreover the enhancement of $\kappa_{xy}^{\rm 2D}/T$ with $H$ from the half-quantized value at around $\mu_0H_{\perp}$=4.9\,T (Figs.\,3b and c) is not reproduced by the numerical calculation based on the perturbation theory.    These discrepancies 
may be attributed to high-field effects and/or non-Kitaev interactions, which deserves further study.

The nearly vanishing of $\kappa_{xy}^{\rm 2D}/T$  after the rapid suppression in the high-field regime (Figs.\,3a, b and c) demonstrates the disappearance of chiral Majorana edge currents.  As shown by the open blue square in Fig.\,1b, the temperature at which $\kappa_{xy}^{\rm 2D}/T$ vanishes  decreases rapidly with decreasing $\mu_0H_{\parallel}$.    This suggests a topological quantum phase transition from the non-trivial  QSL to trivial high-field state, where the thermal Hall effect is absent,   at $\mu_0H_{\parallel}\sim$ 8.5\,T as shown bt the red circle in Fig.\,1c \cite{Jiang}.     The high-field state is likely to be attributed to a non-topological spin liquid phase or a forced ferromagnetic state where the system is nearly fully polarized.  The observation of half-integer thermal Hall conductance reveals that  topologically protected chiral Majorana edge currents persist in $\alpha$-RuCl$_3$,  even in the presence of non-Kitaev interactions and parallel field.   The observation opens a possibility to link  to non-Abelian anyons important for the topological quantum computing, revealing novel aspects of  strongly correlated topological quantum matters.

\section*{Acknowledgements}
We thank S. Fujimoto, H. Ishizuka, N. Kawakami, H.-Y. Kee, E.-G. Moon, M. Shimozawa, K. Sugii,  M. Udagawa, and M. Yamashita for useful discussions. This work was supported by Grants-in-Aid for Scientific Research (KAKENHI) (No.\,25220710, 15H02014, 15H02106, and 15H05457) and Grants-in-Aid for Scientific Research on innovative areas ``Topological Materials Science" (No.\,JP15H05852) from Japan Society for the Promotion of Science (JSPS), and by Yamada Science Foundation and Toray Science Foundation.

%  \section*{Acknowledgements}
% We thank S. Fujimoto, H. Ishizuka, N. Kawakami, H.-Y. Kee, E.-G. Moon, M. Shimozawa, K. Sugii,  M. Udagawa, and M. Yamashita for useful discussions. This work was supported by Grants-in-Aid for Scientific Research (KAKENHI) (No.\,25220710, 15H02014, 15H02106, and 15H05457) and Grants-in-Aid for Scientific Research on innovative areas ``Topological Materials Science" (No.\,JP15H05852) from Japan Society for the Promotion of Science (JSPS), and by Yamada Science Foundation and Toray Science Foundation. 
 
% \section*{Author contributions}
% Y.K. and Y.Matsuda conceived and designed the study. Y.K. and T.O. performed the thermal transport measurements.  N.K. and H.T. synthesized the high-quality single crystalline samples. Y.K., T.O., K.S., M.Y., J.N., Y.Motome, T.S., and Y.Matsuda discussed the results. Y.K., J.N., Y.Motome, T.S., and Y.Matsuda prepared the manuscript. 

% \section*{Author information}
% Reprints and permissions information is available at www.nature.com/reprints. The authors declare no competing financial interests. Readers are welcome to comment on the online version of the paper. Correspondence and requests for materials should be addressed to Y.Matsuda (matsuda@scphys.kyoto-u.ac.jp). 

% \newpage
\clearpage
  \section*{Methods}
 \noindent
 {\bf Single crystal growth.} 
High-quality single crystals of $\alpha$-RuCl$_3$ were grown by a vertical Bridgman method as described in ref.\,\cite{Kubota15}. %The crystal used for thermal transport measurements is identical to sample 1 in ref.\,\cite{Kasahara17}.  %Fine-grained RuCl$_3$ was dehydrated in a quartz tube at 100$^\circ$C for three days. The temperature of the centre of the furnace was set at 1100$^\circ$C and the quartz tube was moved downward in the furnace at a rate of 3 mm/h over 80 h.  
For thermal transport measurements, we carefully picked up thin crystals with plate-like shape.  Typical sample size is  $\sim2$\,mm$\times0.5$\,mm$\times0.02$\,mm. 
We selected the best crystal in which no anomaly associated with the magnetic transition at 14\,K due to the stacking faults is detected by magnetic susceptibility, specific heat, and thermal transport measurements. %The crystal quality has been checked by susceptibility measurements (Extended Data Fig.\,1). 
 
 \noindent
 {\bf Thermal transport measurements.} %For thermal transport measurements, we carefully picked up thin crystals with plate-like shape.  Typical sample size is  $\sim2$\,mm$\times0.5$\,mm$\times0.02$\,mm. 
Thermal and thermal Hall conductivities were measured simultaneously on the same crystal by the standard steady state method, using the experimental setup illustrated in the inset of Fig.\,2a.  Heat current $\bm{q}$ were applied along the $a$ axis ($\bm{q}\parallel\bm{x}$). Magnetic field $\bm{H}$ is applied along various directions in the $ac$ plane (inset of Fig.\,2a). The temperature gradient $-\nabla_xT\parallel\bm{x}$ and $-\nabla_yT\parallel\bm{y}$ was measured by carefully calibrated Cernox thermometers. Sample temperature was measured with accuracy within 0.1\,mK using alternating current resistance bridges.   1\,k$\Omega$ chip resistor was used  to generate the heat current. The magnitude of thermal gradient is less than 5\,\% of the base temperature. 
To reduce the noise level, all measurements are performed in the radio-frequency shielded room. 
For the measurements of the thermal Hall effect, we removed the longitudinal response due to misalignment of the contacts by anti-symmetrizing the measured $\nabla_yT$ as $\nabla_yT^\mathrm{asym}(H)=[\nabla_yT(H)-\nabla_yT(-H)]/2$ at each temperature. 
We note that the offset transverse thermal gradient due to the misalignment of Hall contact was reduced to be less than 0.5\% of longitudinal thermal gradient in zero field.  
$\kappa_{xx}$ and $\kappa_{xy}$ were obtained from longitudinal thermal resistivity $w_{xx}=\nabla_xT/q$ and thermal Hall resistivity $w_{xy}=\nabla_yT^\mathrm{asym}/q$ as $\kappa_{xx}=w_{xx}/(w_{xx}^2+w_{xy}^2)$ and $\kappa_{xy}=w_{xy}/(w_{xx}^2+w_{xy}^2)$. 
To avoid background Hall signal, a LiF heat bath and non-metallic grease were used. We confirmed that the thermal Hall signal in LiF is negligibly small within our experimental resolution \cite{Watanabe16}. 
%The thermal transport measurements were performed on two different single crystals in the same batch,  sample 1 and 2, at Kyoto University and at ISSP, Univ. of Tokyo, respectively, by using setup with different calibrated thermometers.  As shown in Fig.\,2a and 4a, both of $\kappa_{xx}(T)$ and $\kappa_{xy}(T)$ of samples 1 and 2 well  coincide. 

 \newpage

\renewcommand{\figurename}{\textbf{Extended Data Figure}}

\setcounter{figure}{0}

 \begin{figure}[t]
	\begin{center}
		%\vspace*{-20mm}
		%\hspace*{5mm}
		\includegraphics[width=1.0\linewidth]{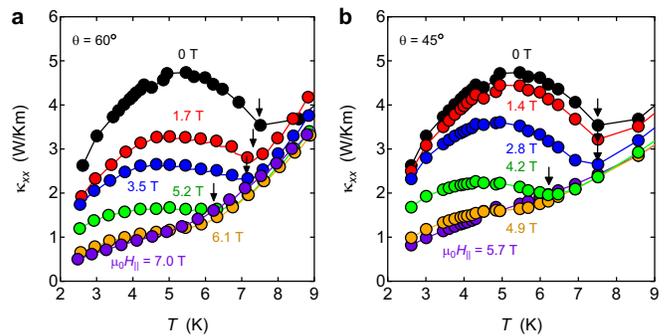}
		%\vspace{10mm}
		\caption{
			{\bf Temperature dependence of the longitudinal thermal conductivity.} 
$\kappa_{xx}$ in tilted field of $\theta=60^\circ$ ({\bf a}) and $45^\circ$ ({\bf b}) plotted as a function of temperature. %Here $\theta$ denotes the angle between the applied magnetic field {\boldmath $H$} and the $c$ axis, and $H_\parallel=\sin\theta$ is the field component parallel to the $a$ axis.   
Arrows indicate the onset temperature of AFM order $T_N$. }
	\end{center}
	%	\vspace{-5mm}
\end{figure}

 \begin{figure}[t]
	\begin{center}
		%\vspace*{-20mm}
		%\hspace*{5mm}
		\includegraphics[width=1.0\linewidth]{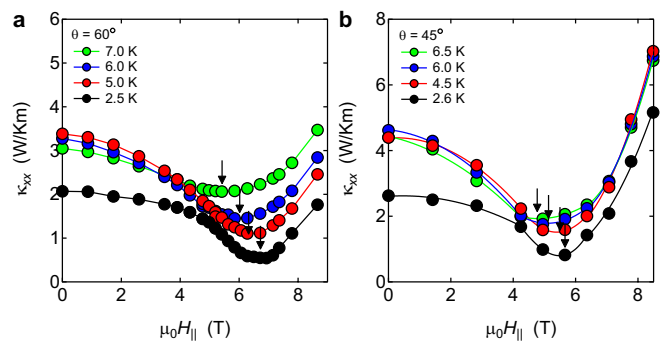}
		%\vspace{10mm}
		\caption{
			{\bf Field dependence of the longitudinal thermal conductivity.} 
$\kappa_{xx}$ in tilted field of $\theta=60^\circ$ ({\bf a}) and $45^\circ$ ({\bf b}) plotted as a function of parallel field component $H_{\parallel}$. Arrows indicate the minimum of $\kappa_{xx}$, which is attributed to the onset field of AFM order. }
	\end{center}
	%	\vspace{-5mm}
\end{figure}

 \begin{figure}[t]
	\begin{center}
		%\vspace*{-20mm}
		%\hspace*{5mm}
		\includegraphics[width=0.85\linewidth]{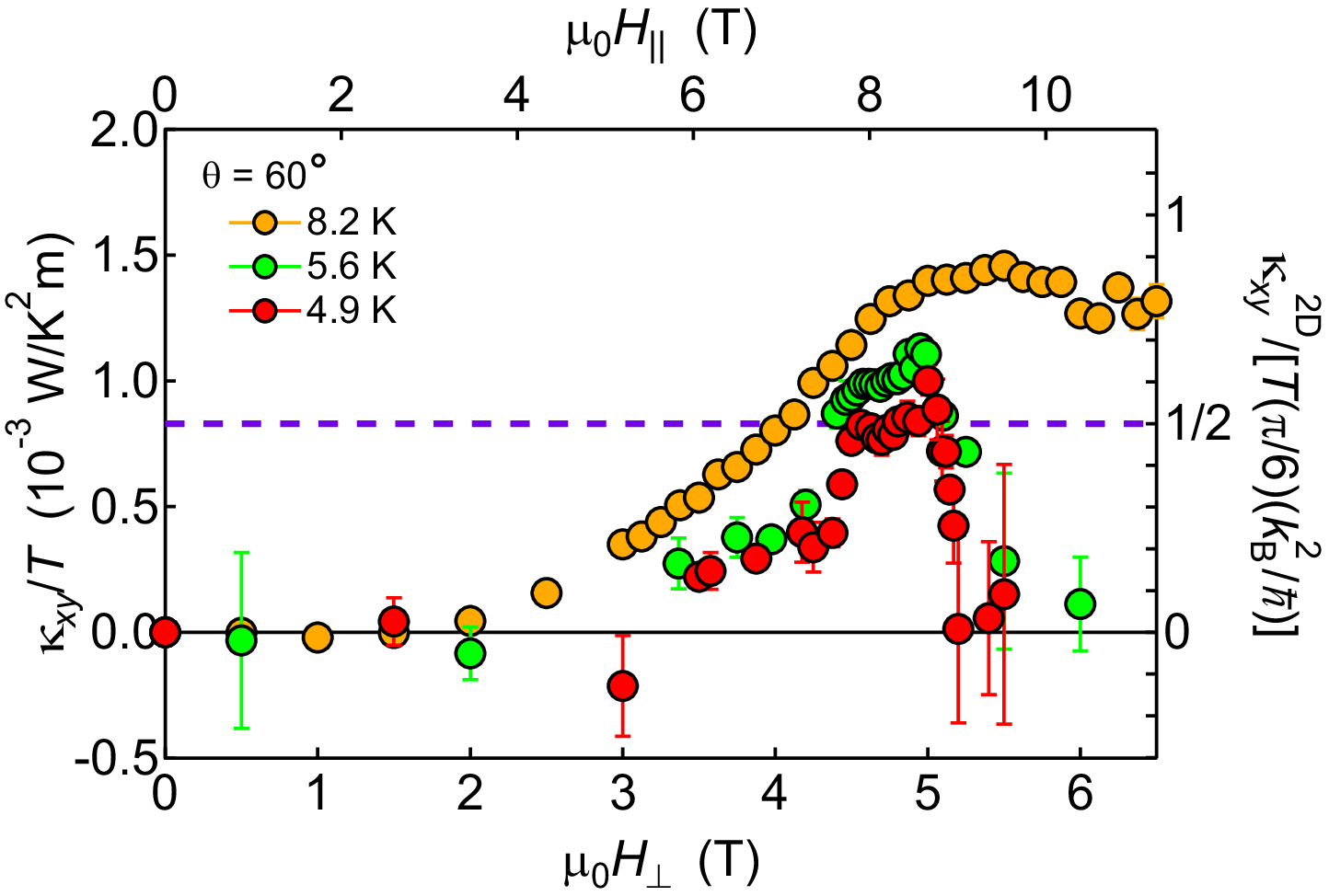}
		%\vspace{10mm}
		\caption{
			{\bf Thermal Hall conductivity in tilted field of $\theta=60^\circ$.} 
Thermal Hall conductivity $\kappa_{xy}/T$ in tilted field of $\theta=60^\circ$ plotted as a function of $H_\perp$. The top axes show the parallel field component $H_\parallel$. The right scales represent the 2D thermal Hall conductance $\kappa_{xy}^{\rm 2D}/T$ in units of $(\pi/6)(k_B^2/\hbar)$. Violet dashed lines represent the half-integer thermal Hall conductance, $\kappa_{xy}^{\rm 2D}/[T(\pi/6)(k_B^2/\hbar)]=1/2$. }
	\end{center}
	%	\vspace{-5mm}
\end{figure}

\clearpage

 \begin{figure}[t]
	\begin{center}
		%\vspace*{-20mm}
		%\hspace*{5mm}
		\includegraphics[width=0.85\linewidth]{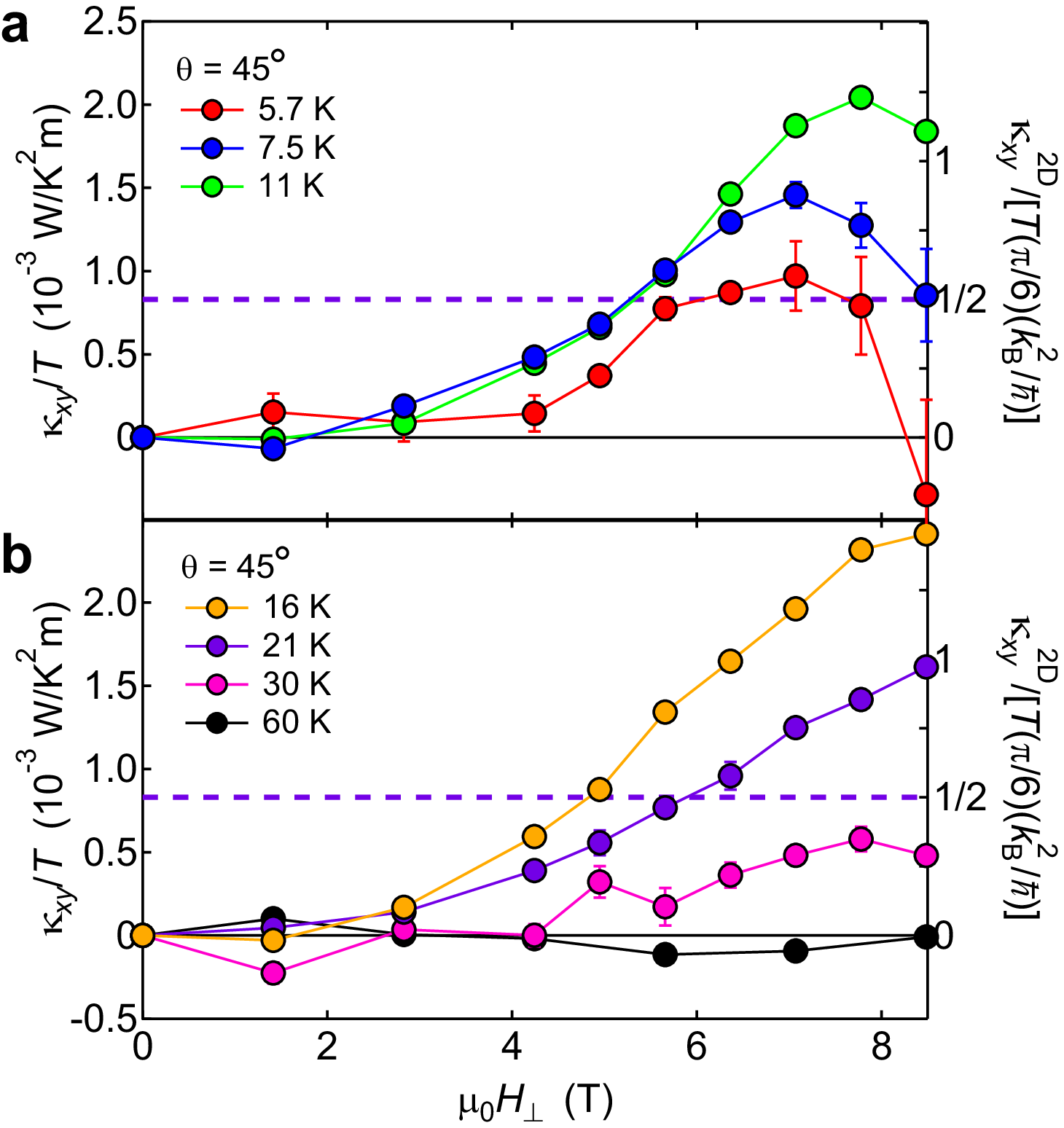}
		%\vspace{10mm}
		\caption{
			{\bf Thermal Hall conductivity in tilted field of $\theta=45^\circ$.} 
{\bf a,b,} Thermal Hall conductivity $\kappa_{xy}/T$ in tilted field of $\theta=45^\circ$ plotted as a function of $H_\perp$. The right scales represent the 2D thermal Hall conductance $\kappa_{xy}^{\rm 2D}/T$ in units of $(\pi/6)(k_B^2/\hbar)$. Violet dashed lines represent the half-integer thermal Hall conductance, $\kappa_{xy}^{\rm 2D}/[T(\pi/6)(k_B^2/\hbar)]=1/2$.}
	\end{center}
	%	\vspace{-5mm}
\end{figure}


\begin{thebibliography}{99}

\bibitem{Kitaev06}
Kitaev, A. Anyons in a exactly solved model and beyond. {\it Ann. Phys.} {\bf 321}, 2-111 (2006). 

%\bibitem{Savary17} Savary, L. \& Balents, L. Quantum spin liquids: a review. {\it Rep. Prog. Phys.} {\bf 80}, 016502 (2017). 


\bibitem{Jackeli09}
Jackeli, G. \& Khaliullin, G. Mott Insulators in the Strong Spin-Orbit Coupling Limit: From Heisenberg to a Quantum Compass and Kitaev Models. {\it Phys. Rev. Lett.} {\bf 102}, 017205 (2009). 

\bibitem{Trebst17}
Trebst, S. Kitaev Materials. Preprint at http://arXiv.org/cond-mat/1701.07056 (2017). 

\bibitem{Kim15}
Kim, H.-S., Shankar, V. V., Catuneanu, A. \& Kee, H.-Y. Kitaev magnetism in honeycomb RuCl$_3$ with intermediate spin-orbit coupling. {\it Phys. Rev. B} {\bf 91}, 241110(R) (2015). 

\bibitem{Banerjee16}
%Banerjee, A., Bridges, C. A., Yan, J.-Q., Aczel, A. A., Li, L., Stone, M. B., Granoroth, G. E., Lumsden, M. D., Yiu, Y., Knolle, J., Bhattacharjee, S., Kovrizhin, D. L., Moessner, R.,, Tennant, D. A., Mandrus, D. G., \& Nagler, S. E., 
Banerjee, A. {\it et al.} Proximate Kitaev quantum spin liquid behaviour in a honeycomb magnet. {\it Nat. Mat.} {\bf 15}, 733-740 (2016). 

\bibitem{Sandilands15}
Sandilands, L. J., Tian, Y., Plumb, W., Kim, Y.-J. \& Burch, K. S. Scattering Continuum and Possible Fractionalized Excitations in $\alpha$-RuCl$_3$. {\it Phys. Rev. Lett.} {\bf 114}, 147201 (2015). 

\bibitem{Nasu16} Nasu, J., Knolle, J., Kovrizhin, D. L., Motome, Y. \& Moessner, R. Fermionic response from fractionalization in an insulating two-dimensional magnet. {\it Nat. Phys.} {\bf 12}, 912-915 (2016).

\bibitem{Yadav16} Yadav, R., Bogdanov, N. A., Katukuri, V. M., Nishimoto, S., van der Brink, J., \& Hozoi, L., Kitaev exchange and field-induced quantum spin-liquid states in honeycomb $\alpha$-RuCl$_3$. {\it Sci. Rep.} {\bf 6}, 37925 (2016).  

\bibitem{Baek} Baek, S.-H.  {\it et al.} Evidence for a Field-Induced Quantum Spin Liquid in $\alpha$-RuCl$_3$. {\it Phys. Rev. Lett.} {\bf 119}, 037201 (2017). 

\bibitem{Wolter} Wolter, A. U. B. {\it et al.} Field-induced quantum criticality in the Kitaev system $\alpha$-RuCl$_3$. {\it Phys. Rev. B} {\bf 96}, 041405(R) (2017). 

\bibitem{Leahy16}
Leahy, I. A., Pocs, C. A., Siegfried, P. E., Graf, D., Do, S.-H., Choi, K.-Y., Normand, B. \& Lee, Minhyea. 
%Leahy, I. A. {\it et al.} 
Anomalous Thermal Conductivity and Magnetic Torque Response in the Honeycomb Magnet $\alpha$-RuCl$_3$. {\it Phys. Rev. Lett.} {\bf 118}, 187203 (2017). 

\bibitem{Hentrich} Hentrich, R. {\it et al.}  Large field-induced gap of Kitaev-Heisenberg paramagnons in $\alpha$-RuCl$_3$, Preprint at http://arXiv.org/cond-mat/1703.08623 (2017). 

\bibitem{Read00}
Read, N. \& Green, D. Paired states of fermions in two dimensions with breaking of parity and time-reversal symmetries and the fractional quantum Hall effect. {\it Phys. Rev. B} {\bf 61}, 10267-10297 (2000). 

\bibitem{Sumiyoshi13}
Sumiyoshi, H. \& Fujimoto, S. Quantum Thermal Hall Effect in a Time-Reversal-Symmetry-Broken Topological Superconductor in Two Dimensions: Approach from Bulk Calculations. {\it J. Phys. Soc. Jpn.} {\bf 82}, 023602 (2013). 

%\bibitem{Nakai17}
%Nakai, R., Ryu, S., \& Nomura, K. Laughlin's argument for the quantized thermal Hall effect. {\it Phys. Rev. B} {\bf 95}, 165405 (2017).
\bibitem{Nomura12}
Nomura, K., Ryu, S.,  Furusaki, A. \&Nagaosa, N.  Cross-Correlated Responses of Topological Superconductors and Superfluids, {\it Phys. Rev. Lett.} {\bf 108}, 026802 (2012).

\bibitem{Nasu17}
Nasu, J., Yoshitake, J. \& Motome, Y. Thermal Transport in the Kitaev Model. {\it Phys. Rev. Lett.} {\bf 119}, 127204 (2017). 

%\bibitem{Winter17}
%Winter, S. M., Riedl, K., Honecker, A. \& Valent\'{i}, R. Breakdown of Magnons in a Strongly Spin-Orbital Coupled Magnet. Preprint at http://arXiv.org/cond-mat/1702.08466 (2017).

\bibitem{MBanerjee17} Banerjee, M., Heiblum, M., Rosenblatt, A., Oreg, Y., Feldman, D. E., Stern, A. \& Umansky, V. Observed quantization of anyonic heat flow. {\it Nature} {\bf 545}, 75-79 (2017). 

\bibitem{Mourik} Mourik, V., Zuo, K., Frolov, S. M., Plissard, S. R., Bakkers, E. P. A. M. \& Kouwenhoven, L. P. Signatures of Majorana Fermions in Hybrid superconductor-Semiconductor Nanowire Devices. {\it Science} {\bf 336}, 1003-1007 (2012). 

\bibitem{Nadj} Nadj-Perge, S., Drozdov, I. K., Li, J., Chen, H., Jeon, S., Seo, J., MacDonald, A. H., Bernevig, B. A. \& Yazdani, A. Observation of Majorana fermions in ferromagnetic atomic chains on a superconductor. {\it Science} {\bf 346}, 602-607 (2014). 

\bibitem{Das12} Das, A., Ronen, Y., Most, Y., Oreg, Y., Heiblum, M. \& Shtrikman, H. Zero-bias peaks and splitting in an Al-InAs nanowire topological superconductor as a signeture of Majorana fermions. {\it Nat. Phys.} {\bf 8}, 887-895 (2012). 

\bibitem{He17} He, Q. L. {\it et al.} Chiral Majorana fermion modes in a quantum anomalous Hall insulator-superconductor structure. {\it Science} {\bf 357}, 294-299 (2017). 
 	
\bibitem{Johnson15} Johnson, R. D. {\it et al.}  Monoclinic crystal structure of $\alpha$-RuCl$_3$ and the zigzag antiferromagnetic ground state. {\it Phys. Rev. B} {\bf 92}, 234119 (2015).

\bibitem{Kasahara17} Kasahara, Y. {\it et al.} Thermal Hall effect in a Kitaev spin liquid: A possible signature of Majorana chiral edge current. Preprint at http://arXiv.org/cond-mat/1709.102806 (2017). 
 	
\bibitem{Majumder} Majumder, M., Schmidt, M., Rosner, H., Tsirlin, A. A., Yasuoka, H. \& Baenitz, M. Anisotropic Ru$^{3+}$ 4$d^5$ magnetism in the $\alpha$-RuCl$_3$ honeycomb system: Susceptibility, specific heat, and zero-field NMR. {\it Phys. Rev. B} {\bf 91}, 180401(R) (2015).  

\bibitem{Chaloupka16} Chaloupka, L. \& Khaliullin, G. Magnetic anisotropy in the Kitaev model systems Na$_2$IrO$_3$ and RuCl$_3$. {\it Phys. Rev. B} {\bf 94}, 064435 (2016).	

\bibitem{Jansa} Jan\v{s}a, N. {\it et al.} Observation of gapped anyons in the Kitaev honeycomb magnet under a magnetic field. Preprint at http://arXiv.org/cond-mat/1706.08455 (2017). 

\bibitem{Banerjee17} Banergee, A. {\it et al.} Excitations in the field-induced quantum spin liquid state of $\alpha$-RuCl$_3$. Preprint at http://arXiv.org/cond-mat/1706.07003 (2017). 

\bibitem{Watanabe16}
%D. Watanabe, K. Sugii, M. Shimozawa, Y. Suzuki, T. Yajima, H. Ishikawa, Z. Hiroi, T. Shibauchi, Y. Matsuda, \&  M. Yamashita, 
Watanabe, D. {\it et al.} 
Emergence of Nontrivial Magnetic Excitations in a Spin Liquid State of Kagome Volborthite. {\it Proc. Natl. Acad. Sci. USA} {\bf 113}, 8653-8657 (2016). 


%\bibitem{Yoshitake16}
%Yoshitake, J., Nasu, J. \& Motome, Y. Fractional Spin Fluctuations as a Precursor of Quantum Spin Liquids: Majorana Dynamical Mean-Field Study for the Kitaev Model. {\it Phys. Rev. Lett.} {\bf 117}, 157203 (2016). 

%\bibitem{Nasu17PRL} Nasu, J., Kato, Y., Yoshitake, J., Kamiya, Y. \& Motome, Y.,  Spin-Liquid-to-Spin-Liquid Transition in Kitaev Magnets Driven by Fractionalization. {\it Phys. Rev. Lett.} {\bf 118}, 137203 (2017).

% 	\bibitem{Cao16}
%	Cao, H. B. Low-temperature crystal and magnetic structure of $\alpha$-RuCl$_3$. {\it Phys. Rev. B} {\bf 93}, 134423 (2016). 

%\bibitem{Yamashita10}
%Yamashita, M., Nakata, N., Senshu, Y., Nagata, M., Yamamoto, H. M., Kato, R., Shibauchi, T., Matsuda, Y. 
%Yamashita, M. {\it et al.} 
%Highly Mobile Gapless Excitations in a Two-Dimensional Candidate Quantum Spin Liquid. {\it Science} {\bf 328}, 1246-1248 (2010). 

%\bibitem{Katsura10} Katsura, H., Nagaosa, N. \& Lee, P. A., Theory of the Thermal Hall Effect in Quantum Magnets, {\it Phys. Rev. Lett.} {\bf 104}, 066403 (2010).

\bibitem{Han16}
Han, J. H. \& Lee, H. Spin Chirality and Hall-Like Transport Phenomena of Spin Excitations. {\it J. Phys. Soc. Jpn.} {\bf  86}, 011007 (2016).

%\bibitem{Han15}
%Lee, H., Han, J. H. \& Lee, P. A. Thermal Hall effect of spins in a paramagnet. {\it Phys. Rev. B} {\bf 91}, 125413 (2015). 

%\bibitem{Strohm05}
%Strohm, C., Rikken, G. L. J. A. \& Wyder, P. Phenomenological Evidence for the Phonon Hall Effect. {\it Phys. Rev. Lett.} {\bf 95}, 155901 (2005). 

\bibitem{Winter16}
Winter, S. M.,  Li, Y., Jeschke, H.O. \& Valent\'{i}, R. Challenges in design of Kitaev materials: Magnetic interactions from competing energy scales. {\it Phys. Rev. B} {\bf 93}, 214431 (2016).

\bibitem{Sugii17}
Sugii, K. {\it et al.} Thermal Hall effect in a phonon-glass Ba$_3$CuSb$_2$O$_9$. {\it Phys. Rev. Lett.} {\bf 118}, 145902 (2017). 

\bibitem{Hirobe17}
Hirobe, D.,  Sato, M. Shiomi, Y, Tanaka, H, \&Saitoh, E.
Magnetic thermal conductivity far above the N?el temperatures in the Kitaev-magnet candidate $\alpha$-RuCl$_3$.
{\it Phys. Rev. B} {\bf 95}, 241112 (2017).



\bibitem{Yu17} Yu, Y. J., Xu, Y., Ran, K. J., Ni, J. M., Huang, Y. Y., Wen, J. S. \& Li, S. Y. Ultralow-temperature thermal conductivity of the Kitaev honeycomb magnet $\alpha$-RuCl$_3$ across the field-induced phase transition. Preprint at http://arXiv.org/cond-mat/1708.04090 (2017). 

\bibitem{Jiang} 
Jiang, H-C.,   Gu, Z-C.,   Qi, X-L., \& Trebst, S. 
Possible proximity of the Mott insulating iridate Na$_2$IrO$_3$ to a topological phase: Phase diagram of the Heisenberg-Kitaev model in a magnetic field.
{\it Phys. Rev. B} {\bf 83}, 245104 (2011). 
	
\end{thebibliography}

\begin{thebibliography}{99}
%\setcounter{enumiv}{31}
 
\bibitem[35]{Kubota15} Kubota, Y., Tanaka, H., Ono, T., Narumi, Y. \& Kindo, K. Successive magnetic phase transition in $\alpha$-RuCl$_3$: XY-like frustrated magnet on the honeycomb lattice. {\it Phys. Rev. B} {\bf 91}, 094422 (2015).  
 
          
          

 \end{thebibliography}
  \end{document}